\documentclass{article}        
\def\bc{\begin{center}}              \def\ec{\end{center}}
\def\beq{\begin{equation}}           \def\eeq{\end{equation}}

\def\bt{\begin{tabular}}             \def\et{\end{tabular}}
\def\la{\langle}    \def\ra{\rangle}  \def\nn{\nonumber}
\def\dg{\dagger}    \def\ci{\cite}   \def\lb{\label}  
        \def\pr{\prime}  
    \def\vs{\vspace}  \def\sm{\small}
   
             \def\lrar{\longrightarrow}
\def\noi{\noindent}
     \def\fns{\footnotesize}

\def\alf{\alpha}     
     
\def\Dlt{\Delta}     \def\dlt{\delta}
\def\lam{\lambda}    \def\Lam{\Lambda}
       \def\sig{\sigma}
\def\z{\zeta}        \def\vphi{\varphi}
   
\def\ome{\omega}      

\begin{document}

\title{\bf Schr\"odinger uncertainty relation and its minimization states
\footnote{Published in journal "the World of
Physics" (Sofia), {\bf24}, No. 2, 107-116 (2001) on the occasion of the 70th
anniversary of the publication of E. Schr\"odinger original paper \ci{S1}.
 Minor improvements are made in this English-translated version.}}
\author{D.A. Trifonov}

\maketitle
physics/0105035

\begin{abstract}
An introductory survey on the Schr\"odinger uncertainty relation and
its minimization states is presented with minimal number of formulas
and some historical points. The case of the two canonical observables, 
position and momentum, is discussed in greater detail: basic
properties of the two subsets of minimization states (canonical
squeezed and coherent states) are reviewed and compared. 
The case of two non-canonical observables is breafly outlined. 
Stanfard SU(1,1) and SU(2) group-related coherent states can be
defined as states that minimize Schr\"odinger inequality for the three
pairs of generators simultaneously. The symmetry of the Heisenberg
and Schr\"odinger relations is also discussed, and two natural
generalizations to the cases of several observables and several
states are noted.
\end{abstract}

\section{The Heisenberg uncertainty principle}

The uncertainty (indeterminacy) principle is a basic feature of quantum
physics. It reveals the fundamental difference between quantum and classical 
descriptions of Nature.
The indeterminacy principle was introduced in 1927 by  Heisenberg \ci{H}
who demonstrated the impossibility of simultaneous precise measurement of
the canonical quantum observables $\hat{x}$ and $\hat{p}_x$ (the
coordinate and the momentum) by positing an approximate relation $\dlt
p_x\dlt x \sim \hbar$,  where $\hbar$ is the Plank constant, $\hbar =
1.05\times10^{-27}$ erg.sec. {\it "The more precisely is the position
determined, the less precisely is the momentum known, and vice versa"}
\ci{H}.
Heisenberg considered this inequality as the "direct descriptive
interpretation"
of the canonical commutation relation between the operators of the
coordinate and momentum:
$[\hat{x},\hat{p}_x] = i\hbar$, $[\hat{x},\hat{p}_x] \equiv
\hat{x}\hat{p}_x-\hat{p}_x\hat{x}$. Qualitative statements about the
repugnance of the precise determination of the coordinate with that of the
momentum have been formulated in 1926 also by Dirac and Jordan ({\fns see
refs. in M.Jammer, {\it The conceptual development of quantum mechanics}, Mc
Graw-Hill, 1967}).  Let us recall that in quantum physics a physical
quantity (observable) $X$ is represented by a Hermitian operator $\hat{X}$
in the Hilbert space of states.  Soon after the Heisenberg paper \ci{H}
appeared  Kennard and Weyl \ci{KW} proved the inequality
\beq\lb{HWK} 
(\Dlt p_x)^2 (\Dlt x)^2 \,\geq\, \hbar/4,
\eeq
where $(\Dlt p_x)^2$ and $(\Dlt x)^2$ are the variances
(dispersions) of $\hat{p}_x$ and $\hat{x}$, defined by Weyl for every
quantum state $|\psi\ra$ via the formula
{\sm $(\Dlt p_x)^2:=
\la\psi|(\hat{p}_x-\la\psi|\hat{p}_x|\psi\ra)^2|\psi\ra,$}
and similarly is  $(\Dlt\hat{x})^2$ defined.
The matrix element {\sm $\la\psi|\hat{X}|\psi\ra\equiv \la \hat{X}\ra$} is
the mean value of the observable {\sm $\hat{X}$} in the state $|\psi\ra$.
The square-root $\Dlt X= \sqrt{(\Dlt\hat{X})^2}$ is called
standard deviation.

In correspondence with the classical probability theory the standard
deviation {\sm $\Dlt X$}  is considered as a measure for the indeterminacy
(or for the fluctuations) of the observable {\sm $\hat{X}$} in the
corresponding state $|\psi\ra$.  In 1930 Dichburn [{\fns R.Dichburn, Proc.
Royal Irish Acad.  {\bf39}, 73 (1930)}] established the relation between
Weyl' $\Dlt x$ and Heisenberg' $\dlt x$ [namely $\Dlt x = \dlt
x/\sqrt{2}$], and proved that the equality $\dlt p_x\dlt x = h/2\pi$ can
be achieved for Gauss probability distributions only. They are these
distributions for which Heisenberg derived his relation, resorting to the
properties of the Fourier transformation.

The Heisenberg-Weyl-Kennard inequality (\ref{HWK}) became known as
the {\it Heisenberg uncertainty relation}.

Generalization of (\ref{HWK}) to the case of arbitrary two observables
(Hermitian operators {\sm $\hat{X}$ and $\hat{Y}$}) was made by Robertson
in 1929 \ci{R1},
\beq\lb{RUR1}
(\Dlt X)^2(\Dlt Y)^2 \,\,\geq\,\,
\frac{1}{4}\left|\la [\hat{X},\hat{Y}]\ra\right|^2.
\eeq
Robertson inequality (\ref{RUR1}) became known again as the Heisenberg
uncertainty relation for two observables {\sm $\hat{X}$ and $\hat{Y}$},
and it is regarded as a mathematical formulation of the Heisenberg
indeterminacy principle for two quantum observables. In view of
this inertia and of the significant Robertson' contribution we shall
refer to the relation (\ref{RUR1}) as Heisenberg--Robertson inequality or
Heisenberg--Robertson uncertainty relation (while (\ref{HWK}) is referred
to as Heisenberg relation).

\section{The Schr\"odinger inequality}

The Heisenberg--Robertson uncertainty relation (\ref{RUR1}) and/or
its particular case (\ref{HWK}) became an irrevocable part of almost every
textbook in quantum mechanics.
However from the classical probability theory it is known that for two random
quantities one can define three second moments: the variances of each
observable and their {\it covariance}. In the relations (\ref{RUR1}) and
(\ref{HWK}) the two variances $\hat{X}$ and $\hat{Y}$ are involved only.
This fact was first noted
by Schr\"odinger in 1930 \ci{S1}, who derived (using Schwartz inequality)
the more general inequality
\beq\lb{SUR}
(\Dlt X)^2(\Dlt Y)^2 - (\Dlt XY)^2 \,\,\geq\,\,
\frac{1}{4}\left|\la [\hat{X},\hat{Y}]\ra\right|^2,
\eeq
where $\Dlt XY$ denotes the covariance \footnote{
Other notations, used in the literature, for the covariance\, (variance),\, 
are\,   
Cov$(X,Y)$,\, $\sig_{XY}$,\, $\Dlt(X,Y)$\,\, (Var$(X)$,\, $\sig_{X}^2$, 
$\Dlt X^2$, $\Dlt^2 X$). 
In his original paper \ci{S1} Schr\"odinger didn't introduce
any symbol for the quantity $(1/2)\la \hat{X}\hat{Y}+\hat{Y}\hat{X}\ra -
\la\hat{X}\ra\la\hat{Y}\ra$, while for the variance (mean-square
deviation) he used $(\Dlt X)^2$.} 
of $\hat{X}$ and $\hat{Y}$,\,\,\,  
$\Dlt XY := (1/2)\la \hat{X}\hat{Y}+\hat{Y}\hat{X}\ra - \la
\hat{X}\ra\la\hat{Y}\ra$.
The ratio $r = \Dlt XY/\Dlt X\Dlt Y$ is called correlation
coefficient for two observables. 
In the classical probability theory the vanishing covariance
is a necessary (but not sufficient) condition for the statistical
independence of two random quantities. Nonvanishing covariance means
stronger correlation between physical quantities.

In the case of coordinate and momentum observables relation (\ref{SUR})
takes the shorter  form of
$(\Dlt x)^2(\Dlt p_x)^2 - (\Dlt xp_x)^2 \,\,\geq\,\, \hbar^2/4$.
For the sake of brevity henceforth we shall work with dimensionless
observables {\sm $\hat{q}=\hat{x}\sqrt{m\ome/\hbar}$} and
{\sm $\hat{p}=\hat{p}_x/\sqrt{m\ome\hbar}$} (instead of $\hat{x}$ and
$\hat{p}_x$), where $m$ and $\ome$ are parameters with dimension of mass
and frequency respectively. For $\hat{q}$ and $\hat{p}$ the Heisenberg and
Schr\"odinger inequalities read simpler: $(\Dlt q)^2(\Dlt p)^2
\,\,\geq\,\, 1/4$,
\beq\lb{qpSUR}
(\Dlt q)^2(\Dlt p)^2 - (\Dlt qp)^2 \,\,\geq\,\, \frac{1}{4}.
\eeq

The Schr\"odinger inequality (\ref{SUR}) is more general and more precise
than that of Heisenberg--Robertson, eq. (\ref{RUR1}): the former is
reduced to the latter in states with vanishing covariance of {\sm $\hat{X}$
and $\hat{Y}$}, and the equality in (\ref{RUR1}) means the equality in
(\ref{SUR}), the inverse being not true. Thus the Schr\"odinger inequality
provides a more stringent limitation (from below) to the product ot two
variances.  Despite of these  advantages the relation (\ref{qpSUR})
and/or (\ref{SUR}) are lacking in almost all text books.  The interest in
Schr\"odinger relation has been renewed in the last two decades only (for
the first time, to the best of my knowledge, in ref. \ci{DKM} -- 50 years
after its discovery) in connection with the description and experimental
realization of the {\it squeezed states} ({\sm SS}) of the electromagnetic
radiation \ci{Caves,obzori}.

Another useful property of the Schr\"odinger uncertainty relation consists
in its form-invariance against nondegenerate linear transformations of the
two observables  \ci{T94}.  If one transforms {\sm $\hat{X},\,\hat{Y}$}
to  {\sm $\hat{X}^\pr,\,\hat{Y}^\pr$},
$$
\hat{X}^\pr =  \lam_{11}\hat{X} + \lam_{12}\hat{Y},\qquad
\hat{Y}^\pr = \lam_{21}\hat{X} + \lam_{22}Y,
$$
one would obtain that the left and the right hand sides of (\ref{SUR}) are
transformed in the same manner ({\it covariantly}) -- by multiplication by
the factor {\sm $(\det\Lam)^2$}, where  $\Lam$ is $2\times 2$ matrix of
the transformation coefficients $\lam_{ij}$. It then follows that if a
given state saturates (\ref{SUR}) for {\sm $\hat{X}$} and {\sm $\hat{Y}$},
it saturates (\ref{SUR}) for the transformed observables
{\sm $\hat{X}^\pr,\,\hat{Y}^\pr$} as well.  The covariance of the
Heisenberg--Robertson  inequality, eq. (\ref{RUR1}), is much more
restricted: both its sides are transformed in the same manner under the
linear scale transformations of the observables {\sm $\hat{X}^\pr =
a\hat{X}$, $\hat{Y}^\pr = b\hat{Y}$} {\it only}.  In the case of canonical
observables this restricted covariance  means that the equality in
(\ref{HWK}) is not invariant under linear canonical transformations, in
particular it is not invariant under rotations in the phase plane.

The scale transformations and the rotations form two distinct subgroups of
the group of linear canonical transformations.  It is quite natural then
to look for another uncertainty relation for $q$ and $p$ which is
covariant (the equality being invariant) under the rotations in the phase
space. It turned out that such covariant inequality does exist: it has the
simple form of
\beq\lb{nner}
(\Dlt q)^2 + (\Dlt p)^2 \,\geq\, 1.
\eeq
This inequality is less precise than (\ref{qpSUR}) and (\ref{HWK}) in the
sense that the equality in it entails the equality in both (\ref{qpSUR})
and (\ref{HWK}). The equality in (\ref{nner}) is invariant under rotations
in the phase plane.

The most precise inequality, the Schr\"odinger one, eq. (\ref{qpSUR}), is
most symmetric  -- the equality in (\ref{qpSUR}) is invariant under both
subgroups (rotations and scale transformations) of the group of linear
canonical transformations of $\hat{q}$ and $\hat{p}$.  For two arbitrary
observables (Hermitian operators) the inequality (\ref{nner}) takes the
form {\small $(\Dlt X)^2 + (\Dlt Y)^2 \,\geq\,|\la[\hat{X},\hat{Y}]\ra|$}.

\section{Minimization of the uncertainty relations }

The interest in minimization of uncertainty relations has increased after
the discovery of {\it coherent states} ({\sm CS}) of the electromagnetic
field in 1963 (independently by the american physicists Klauder, Glauber
and Sudarshan --  see refs. in \ci{KS}), and especially after the
discovery of {\it squeezed states} ({\sm SS}) \ci{Caves,obzori}. Next we
consider the basic properties of states, that minimize Heisenberg and
Schr\"odinger inequalities for the canonical pair $\hat{p},\,\hat{q}$
(subsections {\bf A} and {\bf B}) and for some non-canonical observables:
the spin and quasi-spin components (subsection {\bf C}).

States which minimize the Heisenberg-Robertson inequality (\ref{RUR1})
have been called {\it intelligent} ({\fns C.Aragon et al, J.Math.Phys.
{\bf17} (1976) 1963}), while those which minimize the more general
Schr\"odinger inequality (\ref{SUR}) were named {\it correlated} \ci{DKM}
and {\it generalized intelligent} states \ci{T94}. The names Heisenberg
(Schr\"odinger) intelligent states, Heisenberg (Schr\"odinger) minimum
uncertainty states and Heisenberg (Schr\"odinger) optimal uncertainty
states are also used ({\fns see the review papers D.A.Trifonov, JOSA A {\bf17},
2486 (2000) (e-print quant-ph/0012072); e-print quant/9912084 and refs.
therein}). \\

{\bf A. Minimization of the Heisenberg inequality }\\[-3mm]

As it could be seen from (\ref{SUR}) and (\ref{RUR1}) the problem of
minimization of the Heisenberg-Robertson relation is a particular case of
the minimization of the Schr\"odinger inequality, corresponding to
vanishing correlation coefficient of the two observables.
In fact it was Heisenberg \ci{H} who has first minimized
the inequality (\ref{HWK}), showing that for Gaussian distribution of
$x$ of the form
$\exp[-(x-x^\pr)^2/(\dlt x)^2]$ the equality
$\dlt x\dlt p_x = h/2$ holds.
The problem of minimization of inequalities (\ref{RUR1}), (\ref{qpSUR}) or
(\ref{SUR}) {\it was not considered} in Robertson and Schr\"odinger papers
\ci{R1,S1}.\\

{\small \bf Glauber CS.}
Most widely known minimizing states are the CS  of the electromagnetic
field \ci{KS}, which are considered as states of the {\it ideal
monochromatic laser radiation}. These states (called
Glauber-Klauder-Sudarshan CS, Glauber CS, or canonical CS) are defined as
normalized eigenstates of the non-Hermitian photon annihilation operator
$\hat{a}$, $\hat{a}|\alf\ra =\alf|\alf\ra$, where the complex number
$\alf$ is the eigenvalue of the operator $\hat{a}$. The CS $|\alf\ra$
possess several remarkable physical and mathematical properties, which led
to their intensive applications in many fields of modern theoretical and
mathematical physics.\\[-3mm]

{\fns Before describing the properties of the Glauber CS it would be
useful to recall the {\it stationary states}  $|n\ra$ (with corresponding
wave function $\psi_n(x)$) of the harmonic oscillator. Stationary
states are defined for every quantum system as states with definite energy,
i.e. as eigenstates of the energy operator (the Hamiltonian)
$\hat{H}$. For the mass oscillator (a particle with mass $m$ in the
parabolic potential well $U(x) = m\ome^2 x^2/2$) the Hamiltonian is
$\hat{H} = \hat{p}^2/2m + m\ome^2 x^2/2$. Its eigenvalues (the energy
levels) are discrete and equidistant, $E_n = \hbar\ome(n+1/2)$,
$n=0,1,\ldots$.  The distance between neighbor levels is equal to
$\hbar\ome$.

The transition from $|n\ra$ to $|n\!-\!1\ra$ is performed by the action
of the operator $\hat{a}=(\hat{q}+i\hat{p})/\sqrt{2}$ on $|n\ra$, and the
transition to  $|n\!+\!1\ra$ -- by the action of the conjugated operator
$\hat{a}^\dg = (\hat{q} - i\hat{p})/\sqrt{2}$.
This shows that $\hat{a}$ and $\hat{a}^\dg$ can be regarded as operators,
that annihilate and create photon with Planck energy $\hbar\ome$,
$\hat{a}^\dg\hat{a}$ -- as operator of the number of photons, and $|n\ra$
-- as a state with $n$ photons.  In the context of the electromagnetic
field however $\hat{q}$ and $\hat{p}$ do not have the meaning of coordinate
and moment. For the field in the one-dimensional cavity $\hat{q}$ is
proportional to the electric intensity, and $\hat{p}$ -- to the
magnetic intensity (Loudon and Knight \ci{obzori}).

The time evolution $|n;t\ra$ of an initial state $|n\ra$ is $|n;t\ra =
\exp(-iE_nt/\hbar)|n\ra$. This form shows that the probability
distribution density of the coordinate $|\psi_n(x,t)|^2$ is static:
$|\psi_n(x,t)| = |\psi_n(x)|$.  The energy levels $E_n$ and graphics of
$U(x)$ and $|\psi_1(x)|$ for an oscillator with frequency $\ome\!=\!1/4$
and mass $m\!=\!4$ are shown on figure 1.
In every state $|n\ra$ the mean coordinate and the mean moment are equal
to zero.  The states $|n\ra$ are orthonormalized and form a basis in the
Hilbert space of states, which means that any other state is a
superposition of stationary states}.\\[-2mm]

In Glauber CS $|\alf\ra$ the covariance of the canonical pair of
observables $\hat{p},\,\hat{q}$ vanishes, and the variances of $\hat{q}$ and
$\hat{p}$ are equal: $(\Dlt q)^2 = 1/2,\quad (\Dlt p)^2 = 1/2$. These two
moments minimize the Heisenberg inequality.  Due to this inequality the
value of $1/2$ is the minimal possible one that two dispersions $(\Dlt
p)^2$ and $(\Dlt p)^2$ can take simultaneously. The CS $|\alf\ra$ are the
only states with this property: In any other state at least one of the two
dispersions is greater than $1/2$.  This fact means that in CS $|\alf\ra$
the trajectory of the mass oscillator in the phase space is determined
with the highest possible accuracy. Correspondingly, in the context of
electromagnetic field we have that in CS $|\alf\ra$ the fluctuations of
the electric and magnetic intensities are minimal.
\vspace{10mm}
\bc
\hspace{-1mm}\bt{ll}
\makebox(50,50)[rb]{\begin{minipage}{60mm}
{\small
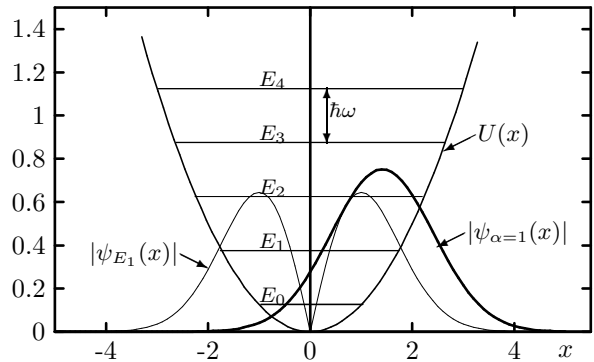
Figure 1. Energy levels $E_n$, potential energy $U(x)$ and absolute values
of wave functions of stationary state $|\psi_{n}(x)|$ with $n\!=\!1$ (two
maximums) and of CS $|\psi_\alf(x)|$ with $\alf=1$ (one Gauss maximum) for
the harmonic oscillator. The graphics of $|\psi_{n}(x)|$ is static, while
the maximum of $|\psi_{\alf}(x)|$ is oscillating harmonically exactly as
the classical particle oscillates.}
\end{minipage}} & \hspace{0mm}\makebox(100,120)[lc]{\input{fig1.pic}}
\et
\ec
\vspace{5mm}

\noi In CS the mean value of the coordinate coincides with the most
probable one and (for the stationary oscillator)
depends on $t$ harmonically, exactly as the coordinate of
classical particle depends on $t$.  In this sense the quantum states
$|\alf\ra$ are {\it "most classical"}.  CS $|\alf\ra$ have the form of an
infinite superposition of $|n\ra$, the mean energy being equal to
$\hbar\ome(|\alf|^2 + 1/2)$.  Graphics of the absolute value of the wave
function $|\psi_\alf(x,t)|$ (the square root of the probability density)
with $\alf = 1$ and $t = 2k\pi/\ome, k=0,1,\ldots$ is shown on figure 1.

CS $|\alf\ra$ possess other "classical properties" as well: minimal energy
of quantum fluctuations, Poisson photon distribution,
{\sm [\,$P_n(\alf) = |\alf|^{2n}\exp(-|\alf|^2)/n!$\,]}, and 
positive Wigner and Glauber-Sudarshan quasi-distributions. The last
property enables one to represent correctly the quantum-mechanical mean
values as classical mean values of the corresponding classical quantities.
However, one can show that in CS $|\alf\ra$ all observables fluctuate,
i.e. these states are not eigenstates of any Hermitian operator.

It is worth noting the physical meaning of the eigenvalue property of CS
$|\alf\ra$ (eigenstates of the ladder operator $\hat{a}$): the
annihilation of one photon in CS doesn't change, up to a normalization
constant, the state of the field. In particular the mean energy (in
$|\alf\ra$ and in the normalized $\hat{a}|\alf\ra$) remains the same.
Destruction of $n$ photons also doesn't change the state, since $|\alf\ra$
is an eigenstate of any power of $\hat{a}$. This remarkable property is
typical for {\it infinite} superpositions of $|n\ra$ {\it only}, as the CS
$|\alf\ra$ are. It is the "infinity" that compensates the
annihilation of the $n$ photons.
Unlike $\hat{a}$, the creation operator $\hat{a}^\dg$ has no eigenstate at
all, and the photon added states $\hat{a}^\dg|\alf\ra$ are no more CS. 
All ladder operators in {\it finite} dimensional Hilbert space have no
eigenstates (except for the state, that is annihilated by the ladder
operator).

\noi
From the remarkable mathematical properties of the canonical CS we will
note here their "overcompleteness" and "orbitality".  The first property
means that the family of CS $|\alf\ra$ is overcomplete in the Hilbert
space, i.e. any other state $|\psi\ra$ can be represented as a continuous
superposition of CS: $|\psi\ra = (1/\pi)\int |\alf\ra\la\alf|\psi\ra
d^2\alf$.  This overcompleteness enables one to represent states as
analytic functions of $\alf$ (or as functions in the phase space
respectively), and abstract operators -- as differential operators.  For
example, $\hat{a} = d/d\alf$ and $\hat{a}^\dg = \alf$.  This {\it CS
representation} provides the possibility to use powerful analytic method
in treating various problems of quantum physics. It is very convenient in
elucidating the relationship between quantum and classical description of
physical systems.

The orbitality property consists in the fact, that the family of CS
$|\alf\ra$ is an orbit of the unitary Weyl displacement operators
$\hat{D}(\alf) = \exp(\alf\hat{a}^\dg-\alf^*\hat{a})$ through the ground
state $|0\ra$, i.e. $|\alf\ra = \hat{D}(\alf)|0\ra$.  One says that
Glauber CS are generated from the vacuum $|0\ra$ by the action of Weyl
operators.  The set of Weyl operators form an unitary representation of
the group of Heisenberg-Weyl.  As early as in 1963 Klauder suggested that
overcomplete families of states could be constructed using unitary
representations of other Lie groups.  For the group of rotations $SO(3)$
such overcomplete family was constructed in 1971 by Radcliffe (spin CS), 
and for the group of pseudo-rotations $SO(1,2)$ -- by Solomon in 1971 ({\fns 
A. Solomon, J.Math.Phys. {\bf12}, 390 (1971)}) and Perelomov in 1972
(quasi-spin CS) (see refs. in \ci{KS}).  Perelomov proved that orbits of
operators of irreducible unitary representations of any Lie group do form
overcomplete sets of states.  The quantum evolution operators of system
with Lie group symmetry are operators of the corresponding unitary
representations of the group. This gives the idea how to generate
physically new families of states starting from known initial ones. \\

{\bf \sm General form of states, minimizing Heisenberg inequality.} Glauber CS
are not the most general ones that minimize the Heisenberg relation
 (\ref{HWK}). Evidently, if in (\ref{HWK}) one variance increases
$\kappa$-times, and the other decreases $\kappa$-times, the equality in
(\ref{HWK}) will be preserved.  Such change in the variances
can be achieved acting on CS $|\alf\ra$ by the unitary operator
$\hat{S}(s) = \exp[s(\hat{a}^{\dg 2} - \hat{a}^2)/2]$, where $s$ is real
parameter \ci{Stoler}.
In states $|\alf,s\ra = \hat{S}(s)|\alf\ra$ we have
$\Dlt p \!=\! e^{-s}\frac{1}{\sqrt{2}}$ and
$\Dlt q \!=\!e^{s}\frac{1}{\sqrt{2}}$. One sees that for $s\neq 0$ one of the
variances is decreased below the value of $1/2$.
States in which $\Dlt q$ or $\Dlt p$ is below
$1/\sqrt{2}$ (the value of $\Dlt q$ and $\Dlt p$ in the CS), have been
called {\it squeezed states}.

States $|\alf,s\ra$ are the most general ones that minimize Heisenberg
inequality.  However these states are {\it extremely unstable} in time
({\fns D.Stoler, Phys.Rev. D{\bf11}, 3033 (1975); D.Trifonov, Phys.Lett.
A{\bf48} (1974) 165}).  If the oscillator is prepared at $t=0$ in a state
$|\alf,s\ra$ with $s\neq 0$, then at $t>0$ it goes out of the family
$\{|\alf,s\ra\}$  and the equality in (\ref{HWK}) is violated. 
Moreover, in the evolved
oscillator state $|\alf,s;t\ra$ inevitably the covariance of $\hat{q}$ and
$\hat{p}$ is generated, which is not taken into account in the Heisenberg
relation. In the Heisenberg picture the free oscillator evolution operator
acts as rotation on angle $-\ome t$ in the phase space. These rotations do
not preserve the initial form of the relation (\ref{HWK}).

Unlike the Heisenberg relation (\ref{HWK}), the inequality (\ref{nner}) is
invariant under rotations in phase plane, and this is another explanation
of  the temporal stability of $|\alf\ra$ under free oscillator evolution.
It is the uncertainty relation (\ref{nner}) that is minimized in CS
$|\alf\ra$ only: in any other state $(\Dlt q)^2 + (\Dlt p)^2 > 1$.\\[3mm]

{\bf B. Minimization of the Schr\"odinger inequality. Squeezed states}\\[-3mm]

The problem of minimization of the Schr\"odinger inequality (\ref{qpSUR})
was first considered (as far as I know) in 1980 in \ci{DKM}, where it was
shown that eigenstates $|\beta,\lam\ra$ of the operator $\lam \hat{q} +
i\hat{p}$, minimize (\ref{qpSUR}) for every complex $\lam$ and $\beta$.  At
Im$\lam = 0$ one has $\Dlt qp = 0$ and the solutions $|\beta,\lam\ra$
coincide with the above $|\alf,s\ra$. For complex $\lam$ the states
$|\beta,\lam\ra$ turned out to coincide \ci{T94} with the Stoler states
$|\alf,\z\ra$ introduced providently in his 1970 paper \ci{Stoler}:
$|\alf,\z\ra = S(\z)|\alf\ra$, where $S(\z) = \exp[(\z\hat{a}^{\dg 2} -
\z^*\hat{a}^2)/2]$, $\z \in \mathbf{C}$.  The family $\{|\alf,\z\ra\}$
received broad popularity as the family of {\it squeezed states} (called
also canonical squeezed states), and the unitary operator $S(\z)$
became known as {\it squeeze operator}) \ci{obzori}.

The canonical SS can be defined \ci{DKM,T94} as states that minimize
Schr\"odinger uncertainty relation (\ref{qpSUR}), i.e. as solutions to
the equation
\beq\lb{|a,u,v>} 
(\mu\hat{a} + \nu\hat{a}^{\dg})\,|\alf,\mu,\nu\ra = \alf|\alf,\mu,\nu\ra.
\eeq
where $\mu$ and $\nu$ are {\it complex} parameters, and $|\mu|^2-|\nu|^2=1$.
It is the equation (\ref{|a,u,v>}) where the alternative notation
(the Yuen notation -- see ref. in \ci{obzori}) $|\alf,\mu,nu\ra$ for
the squeezed states stems from. For Stoler states $|\alf,\z\ra$ one has
$\mu={\rm ch}|\z|$, $\nu= -{\rm sh}|\z|\exp(i\arg\z)$. At $\nu = 0$ the
canonical CS $|\alf\ra$ are reproduced.  The free oscillator evolution of
SS $|\alf,\mu,\nu\ra$ is stable (the equality in (\ref{qpSUR}) is
preserved by any Hamiltonian at most quadratic in $\hat{p},\,\hat{q}$
\ci{T94}). For the electromagnetic field these states are experimentally
realized, the corresponding light being called squeezed \ci{obzori}.  The
SS $|\alf,\mu,\nu\ra$ can be generated by letting the laser light (which
is supposed to be in a Glauber CS $|\alf\ra$) pass through nonlinear
optical media.  The simplest optical nonlinear interaction is described by
a Hamiltonian which is a linear combination of squared $\hat{a}$ and
$\hat{a}^\dg$.  The quantum evolution operator, corresponding to such
quadratic interaction takes exactly the form of the squeeze operator
$S(\z(t))$, i.e.  the evolved CS is of the form of SS $|\alf,\z\ra$.

The wave function $\psi_\alf(x;\mu,\nu)$ of SS $|\alf,\mu,\nu\ra$ is
a quadratic in terms of $x$ exponent, a particular case of which is the
wave function of CS $|\alf\ra$. This implies that some of the properties
of the two kinds of states should be similar.  An examples of such similar
properties are the circular form of the oscillator phase space trajectories
(see Fig. 2), and the coincidence of the mean coordinate and moment with
the most probable ones. Let us recall that in stationary states
the trajectory is degenerated into a point and the mean coordinate and
moment deviate from the most probable ones.
\vspace{15mm}

\bc
\hspace{-10mm}\bt{ll}
\makebox(50,50)[rb]{\begin{minipage}{60mm}
{\small
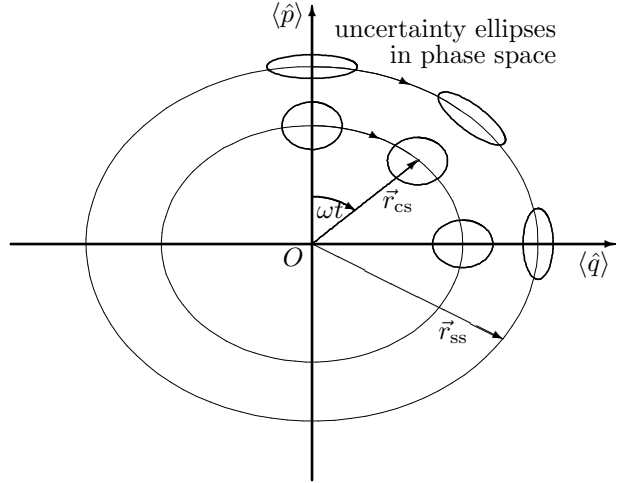
Figure 2. Trajectories of the mean values of $\hat{p}$ and $\hat{q}$ and
uncertainty ellipses in CS $|\alf;t\ra$ (a circle with radius
$r_{cs}$) and in SS $|\alf,\mu,\nu;t\ra$ (a circle with radius $r_{ss}$)
of the free oscillator.  The means are oscillating
with frequency $\ome$, and the variances -- with $2\ome$.  At $\alf=0$
(vacuum and squeezed vacuum) the trajectories are degenerated into a point
($r_{cs} \!=\! 0 \!=\! r_{ss}$).}
\end{minipage}} & \hspace{-10mm}\makebox(100,120)[lc]{\input{fig2xx.pic}}
\et
\ec
\vspace{7mm}
\noi The free oscillator time-evolution preserves Schr\"odinger
intelligent states stable :
$|\alf,\mu,\nu;t\ra = |\alf(t),\mu(t),\nu(t)\ra$,\,\, $\mu(t)=\mu
\exp(i\ome t)$, $\nu(t)=\nu \exp(-i\ome t)$. 
The centers of the wave packets of the
time-evolved states $|\alf(t),\mu(t),\nu(t)\ra$ and CS $|\alf(t)\ra$
oscillates with the same period
$2\pi/\ome$ (while in $|n;t\ra$ the picture is static).
Fluctuations in $\hat{q}$ and $\hat{p}$ in $|\alf(t),\mu(t),\nu(t)\ra$
are oscillating in time with frequency $2\ome$, their sum remaining constant:
$(\Dlt q)^2(t) = |\mu(t)-\nu(t)|^2/2$, $(\Dlt p)^2(t) =
|\mu(t)+\nu(t)|^2/2$. These two variances, the corresponding covariance
$\Dlt qp(t) = {\rm Im}[\mu^*(t)\nu(t)]$ and the mean values
$\la\cdot\ra_t$ of $q$ and $p$ in the evolved state determine an
ellipse in the phase space,
\begin{eqnarray}\lb{ellipse}
(\Dlt p)^2(t)\,(q-\la q\ra_t)^2 + (\Dlt q)^2(t)\, (p-\la p\ra_t)^2 -
2\Dlt qp(t)\,(q-\la q\ra_t)(p-\la p\ra_t) = \nn\\
(\Dlt p)^2(t_0)(\Dlt q)^2(t_0),\nn
\end{eqnarray}
where the initial moment $t_0$ is chosen such that $\Dlt qp(t_0)=0$.  This
ellipse is known as the {\it uncertainty ellipse}.  It is also called 
ellipse of equal probabilities or Wigner ellipse, since Wigner
quasi-distribution of $|\alf,\mu,\nu\ra$ is constant on it.  The
semiaxes of the Wigner ellipse are just the initial standard
deviations $\Dlt q(t_0)$ and $\Dlt p(t_0)$. Note that the current
variances are $(\Dlt q)^2(t) = (\Dlt q)^2(t_0)\cos^2(\ome t)+(\Dlt
p)^2(t_0)\sin^2(\ome t)$, $(\Dlt p)^2(t) = (\Dlt q)^2(t_0)\cos^2(\ome
t)-(\Dlt p)^2(t_0)\sin^2(\ome t)$, and these are not equal to projections 
of the uncertainty ellipse on to the coordinate axes.  

At $t=t_0$ the semiaxes are parallel to the coordinate axes. At $t>t_0$
they are rotated to an angle of $\vphi=-\ome t$.  When $\vphi =
0,\pi,\ldots$ the covariance $\Dlt qp(t)=0$, that is the covariance $\Dlt
qp(t)$ determines the inclination of the ellipse axes to the coordinate
axes.  Thus the free field time-evolution rotates the Wigner ellipse,
preserving the length of its semiaxes (therefore its area is also constant)
(see Fig. 3). The (stationary and nonstationary) oscillator time-evolution
of the variances of $\hat{q}$ and $\hat{p}$ in Gaussian wave packets has
been studied as early as in 1974 ({\fns M.Sargent, M.Scully and W.Lamb, {\it
 Laser physics}, Addison-Wesley, 1974; D.Trifonov, Phys.Lett. A{\bf48}
(1974) 165} ).

Despite of the functional closeness of the wave functions of SS
$|\alf,\mu,\nu\ra$ and CS $|\alf\ra$ some of their physical properties are
significantly different. The main difference consists in the squeeze-effect,
where the name SS for $|\alf,\mu,\nu\ra$ originates from:
at $\nu\lrar \mu$ ($\nu\lrar -\mu$) the fluctuations of $\hat{q}$ (of
$\hat{p}$) in $|\alf,\mu,\nu\ra$ decrease below their value in CS and tend
to zero (ideal squeezing -- ideal SS). Note however that when Re$(uv^*)=0$
there is no squeezing -- both variances are greater than $1/2$.

Another difference, that was intensively discussed in the literature is
the non-positivity of the Glauber-Sudarshan quasi-probability distribution
$P(\alf^\pr)$ (defined by means of $\hat{\rho} = \int P(\alf^\pr)
|\alf^\pr\ra\la\alf^\pr|d^2\alf^\pr$, where $\hat{\rho}$ is pure or mixed
state). For CS $|\alf\ra$ one has $P(\alf^\pr) > 0$, while for
$|\alf,\mu,\nu\ra$ the function $P(\alf^\pr)$ may be negative over some
range of $\alf^\pr$ (see \ci{obzori} and refs. therein). Due to violation
of the positivity of this quasi-probability $|\alf,\mu,\nu\ra$ became
known as {\it non-classical states}.  This non-classicality is closely
related to the squeeze-effect: reduction of the fluctuations of $\hat{q}$
or $\hat{p}$ in a given quantum state is a sufficient condition for the
non-positivity of $P(\alf^\pr)$.  The third difference between CS and SS
is related to their photon distributions. The Poisson photon distribution
in the laser radiation (described by the CS $|\alf\ra$) is considered as a
classical one.  Its main feature is the equality of photon number variance
with the mean number of photons. This equality is violated in states
$|\alf,\mu,\nu\ra$ with  $\nu\neq 0$.  Number distribution with $(\Dlt
n)^2 > \la\hat{n}\ra$ is called super-Poissonian, and with $(\Dlt n)^2 <
\la\hat{n}\ra$ -- sub-Poissonian \ci{obzori}. Examples of sub- and
super-Poissonian distributions in states $|\alf,\mu,\nu\ra$ are
shown on figure 3 (graphics b and c).

Again there is a relation to the non-positivity of the Glauber-Sudarshan
quasi-distribution: the latter is non-positive definite, if $(\Dlt n)^2 <
\la\hat{n}\ra$. Due to this property states with sub-Poissonian statistics
are considered as non-classical.  Experimentally the sub-Poissonian
statistics is revealed as {\it photon antibunching} (impossibility to
detect photons in arbitrary closed moments of time). Ideal photon
antibunching (or maximal non-classicality) is exhibit in the states
$|n\ra$ with definite number of photons, for which $\Dlt n = 0$,
$\la\hat{n}\ra=n$.  In contradistinction  to the Poisson case the
sub- and super-Poisson distributions may strongly oscillate.
\vspace{10mm}

\bc
\hspace{-15mm}\bt{ll}
\makebox(50,50)[rb]{\vs{-3mm}
\begin{minipage}{55mm}
{\small
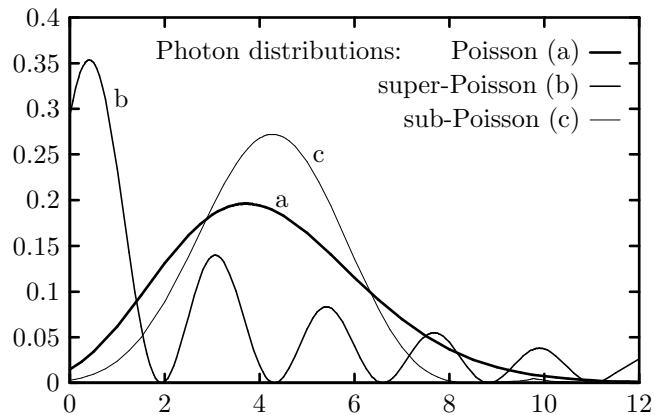
Figure 3. Photon distributions in Schr\"odinger intelligent states
$|\alf,\mu,\nu\ra$ with one and the same mean number of photons $\la
\hat{a}^\dg \hat{a}\ra \simeq 4.22$.  Oscillations are typical to states
with $\nu\neq 0$. In the cases (b) and (c) oscillations occur for large $n$,
$n\rightarrow \infty$, as well, but in (c) they are invisible in the scale
used}.
\end{minipage}} $\quad$& \hspace{-5mm}\makebox(100,120)[lc]
{\input{fig3.pic}}
\et
\ec
\vspace{10mm}

Examples of
oscillating sub- and super-Poisson photon distributions in Schr\"odinger
intelligent states $|\alf,\mu,\nu\ra$ are shown in Fig. 3.\\[3mm]

{\bf C. Minimization states for noncanonical observables}\\[-3mm]

An {\it important application} of the states $|\alf,\mu,\nu\ra$, which
minimize the inequality (\ref{qpSUR}) for canonical pair
$\hat{p}$--$\hat{q}$, was pointed out by Caves in 1981 \ci{Caves}, who
analyzed the accuracy of the interferometric measurements of weak
signals, such as the detection of the gravitational waves.
He found, that the measurement accuracy can be significantly increased if
laser light used in the interferometer is replaced with squeezed light,
described by $|\alf,\mu,\nu\ra$. This fact motivated the search of
squeezed states for other pairs of (noncanonical) observables
{\sm$\hat{X},\,\hat{Y}$}, i.e. of field states with reduced quantum
fluctuations of {\sm$\hat{X}$} or {\sm $\hat{Y}$}.
In paper \ci{T94} it was proposed to construct SS for two general
observables as states, that minimize Schr\"odinger inequality (\ref{SUR}).
Such minimizing states were called \ci{T94} $\hat{X}$--$\hat{Y}$
generalized intelligent states, or $\hat{X}$--$\hat{Y}$ Schr\"odinger
intelligent states. Following \ci{DKM} they could be called
$\hat{X}$--$\hat{Y}$ correlated states. The inequality (\ref{SUR}) is
minimized in a state $|\psi\ra$ if and only if $|\psi\ra$ is an eigenstate
of (generally {\it complex}) combination of {\sm$\hat{X}$} and
{\sm$\hat{Y}$}, i.e. if $|\psi\ra $ satisfies the equation \ci{T94}
\beq\lb{|zuv>} 
(u\hat{A}+v\hat{A}^\dg)|\psi\ra = z|\psi\ra,
\eeq
where $z,\,u,\,v\in \mathbf{C}$, {\sm $\hat{A}= \hat{X}+i\hat{Y}$}.  This
equation shows that at $v\lrar u$ ($v\lrar -u$) the solution
$|\psi\ra\equiv |z,u,v\ra$ must tend to the eigenstate of {\sm$\hat{X}$}
(of {\sm$\hat{Y}$}), i.e. at  $v\lrar u$ ($v\lrar -u$) quantum
fluctuations of {\sm$\hat{X}$} ({\sm$\hat{Y}$}) must tend to zero.
Therefore the solutions $|z,u,v\ra$ to (\ref{|zuv>}), when exist, are
ideal squeezed states for the corresponding two observables (shortly
$\hat{X}$--$\hat{Y}$ SS).

The equation (\ref{|zuv>}) was solved \ci{T94} for the pairs of Hermitian
generators (observables) {\sm$K_1,\, K_2$} and {\sm$J_1,\,J_2$} of the
groups {\sm$SU(1,1)$} and {\sm$SU(2)$}.  It turned out that in these cases
the family of solutions $|z,u,v;k\ra$ ($|z,u,v;j\ra$) contains the
standard {\sm$SU(1,1)$}) and {\sm$SU(2)$} CS \ci{KS}, i.e.  all these
group-related CS minimize the Schr\"odinger uncertainty relation for the
first pair of the group generators.  Generators of {\sm$SU(1,1)$} and
{\sm$SU(2)$} have important boson/photon realizations ({\fns see, e.g. 
quant-ph/9912084 and refs. therein}). For example, the $SU(1,1)$
generators can be realized by means of one pair of boson annihilation and
creation operators $\hat{a}$, $\hat{a}^\dg$ as

$$\hat{K}_1 = \frac 14(\hat{a}^2+\hat{a}^{\dg 2}),\quad
\hat{K}_2 = \frac i4(\hat{a}^2 - \hat{a}^{\dg 2}),\quad
\hat{K}_3 = \frac 14(2\hat{a}^\dg \hat{a} + 1).
$$

States $|z,u,v;k\ra$ for this realization (here $k=1/4, 3/4$)
can exhibit squeezing in fluctuations not only of
{\sm$\hat{K}_1$} and {\sm$\hat{K}_2$}, but also of
$\hat{q}$ and $\hat{p}$. In a certain range of parameters $u$ and $v$
squeezing may occur for {\sm$\hat{K}_1$} and $\hat{p}$ (or for
{\sm$\hat{K}_2$} and $\hat{q}$ simultaneously (joint squeezing for two
noncommuting observables).
Under free field evolution the states $|z,u,v;k\ra$ and $|z,u,v;j\ra$
remain stable (as the canonical SS $|\alf,\mu,\nu\ra$ do).  This means
that if the electromagnetic radiation is prepared in such states, it
would propagate stable in vacuum, and in a linear and homogeneous media
as well.  {\it Schemes for generation} of one-mode and two-mode light in
states of the families $\{|z,u,v;k\ra\}$ and $\{|z,u,v;j\ra\}$ are
proposed in several papers.  Brif and Mann [{\fns C.Brif and A.Mann, Phys.
Rev. A{\bf54}, 4505 (1996)}] showed that, light in states $|z,u,v;k\ra$
(for {\sm$\hat{K}_2$--$\hat{K}_3$}) and in $|z,u,v;j\ra$ (for
{\sm$\hat{J}_2$--$\hat{J}_3$}) could be used for further significant
increase of the accuracy of the interferometric measurements.

Schr\"odinger intelligent states are constructed analytically for every
pair of the quasi-spin ($K_i$--$K_j$) and spin ($J_i$--$J_j$) components
({\fns see e.g. quant-ph/9912084; JOSA A 17 (2000)
2486, and refs.  therein}).  Are there states of systems with $SU(1,1)$
($SU(2)$) symmetry that minimize Schr\"odinger uncertainty relation for
all the three pairs $K_1$--$K_2$, $K_2$--$K_3$ and $K_3$--$K_1$
($J_1$--$J_2$, $J_2$--$J_3$ and $J_3$--$J_1$) {\it simultaneously}\,?
The answer to this question is positive: such states with optimally
balanced fluctuations  of the three observables $K_1,K_2,K_3$
($J_1,J_2,J_3$) (states with maximal $su(1,1)$ or $su(2)$ intelligency)
are the known Klauder-Perelomov $SU(1,1)$ CS (Radcliffe-Gilmore $SU(2)$
CS) {\it only} ({\sm proof in quant-ph/9912084 and
in J. Phys. A 31 (1998) 8041}). For the above noted one-mode
realization of the $su(1,1)$ these group related CS coincide with the
known squeezed vacuum states and in our notations here they are
$|\alf\!=\!0,\mu,\nu\ra$. In the previous subsection we have seen that
$|\alf\!=\!0,\mu,\nu\ra$ minimize the Schr\"odinger relation for $\hat{q}$
and $\hat{p}$ as well. Hence the squeezed vacuum states are the unique
states that minimize Schr\"odinger inequalities for four pairs of
observables $K_i$--$K_j$ and $q$--$p$ simultaneously.  They are these
states that was used by Caves \ci{Caves} to increase the accuracy of the
interferometric measurements.\\

\section{\bf Generalizations of the Schr\"odinger uncertainty relation}

The Heisenberg and Schr\"odinger uncertainty relations reveal
quantitatively the statistical correlations between two observables in one
and the same state.
Two natural questions can be immediately
formulated: are there statistical correlations \,\,
a) between several observables in one state?\,\,
b) between observables in two and more states?\,\,

The positive answer to the first question was given by Robertson in 1934
[{\fns Phys. Rev.  {\bf46} 794 (1934)}]\, by proving of the inequality
\beq\lb{RUR}
\det \sig \geq \det\,C,
\eeq
where $\sig$ is the matrix of all second moments (the uncertainty, the
covariance or dispersion matrix) of $n$ observables
{\sm$\hat{X}_1,\ldots,\hat{X}_n$}, and $C$ is the matrix of mean values of
their commutators, $\sig_{jk} = \la X_jX_k+X_kX_j\ra/2 + \la X_j\ra\la
X_k\ra \equiv \Dlt X_jX_k$, {\sm$C_{jk} = -(i/2)\la[X_j,X_k]\ra$}.  At
$n=2$ Robertson uncertainty relation (\ref{RUR}) coincides with the
Schr\"odinger one, eq. (\ref{SUR}). The minimization of Robertson
inequality (\ref{RUR}) is considered in \ci{T97}.

The second question also has a positive answer\,\, [{\fns D.A.Trifonov, 
J. Phys. A{\bf33} (2000) L296}]. Here is the invariant generalization
of the Schr\"odinger relation (\ref{qpSUR}) for $\hat{p}$ and $\hat{q}$
to the case of two states $|\psi\ra$ and $|\phi\ra$,
\beq\lb{esur}
\mbox{$\frac 12$}\left[(\Dlt_\psi q)^2(\Dlt_\phi p)^2 + (\Dlt_\phi q)^2
(\Dlt_\psi p)^2\right] - \left|\Dlt_\psi qp\,\Dlt_\phi qp\right|\,\,
\geq\,\,
\frac 14\, ,
\eeq
where $(\Dlt_\psi qp)^2$ is the covariance of $\hat{q}$ and $\hat{p}$ in
the state $|\psi\ra$.
At $|\psi\ra=|\phi\ra$ this inequality reproduces that of Schr\"odinger,
eq. (\ref{qpSUR}). The relation (\ref{esur}) is neither a sum nor a
product of the two Schr\"odinger relations for $|\psi\ra$ and $|\phi\ra$
correspondingly. It can not be represented as a sum or as a product of two
quantities, each one depending on one of the two states only. Such
unfactorizable uncertainty relations are called {\it state
entangled}.  The inequality (\ref{esur}), and (\ref{qpSUR}) as well,
contains the second statistical moments of $\hat{q}$ and $\hat{p}$, which
are measurable quantities.  The experimental verification of the relation
(\ref{esur}) would be, we hope, a new confirmation of the Hilbert space
model of quantum physics.

\vspace{5mm}

{\large \bf References, pointed in the Text:}

\begin{enumerate}

\item Max Jammer, {\it The conceptual development of quantum mechanics},
      (Mc Graw-Hill, New York, $\ldots$, Sydney, 1967) (Russian
      Translation:  Nauka, 1985).

\item R.W. Dichburn, {\sm\it The uncertainty principle in quantum
mechanics}, Proc. Royal Irish Acad. {\bf39}, 73 (1930).

\item C. Aragone, E. Chalbaud and S. Salamo, {\sm\it On intelligent spin
      states}, J. Math. Phys. {\bf17}, 1963-1971 (1976).

\item D.A. Trifonov, {\sm\it Generalized uncertainty relations and coherent
      and squeezed states}, JOSA A {\bf17}, 2486 (2000)
	  (e-print quant-ph/0012072).

\item D.A. Trifonov, {\sm\it The uncertainty way of generalization of
      coherent states}, e-print quant-ph/9912084.

\item M. Sargent, M.O. Scully and W.E. Lamb, {\it Laser physics} (Reading:
      Addison-Wesley, 1974).

\item A.I. Solomon, J. Math. Phys. {\bf12}, 390 (1971).

\item D. Stoler, {\sm\it Most general minimality preserving Hamiltonian},
      Phys.Rev. D {\bf11}, 3033 (1975). 

\item D.A. Trifonov, {\sm\it Coherent states and uncertainty relations},
      Phys. Lett. A {\bf48} (1974) 165.

\item C. Brif and A. Mann, {\sm\it Nonclassical interferometry with
      intelligent light}, Phys. Rev. A {\bf54}, 4505 (1996).

\item H.P. Robertson, {\sm\it An indeterminacy relation  for several
      observables and its classical interpretation}, Phys. Rev. {\bf46}
	  794 (1934).

\item D.A. Trifonov, {\sm\it State extended uncertainty relations},
      J. Phys. A {\bf33} (2000) L296.

\end{enumerate}

\end{document}